\documentstyle[pre,aps]{revtex}
\begin{document}
\draft
\title{
Unconventional quasiparticle lifetime in graphite} 
\author{J. Gonz\'alez$^{\dag}$, F. Guinea$^{\ddag}$ and
M. A. H. Vozmediano$^*$ \\}
\address{
$^{\dag}$ 
Instituto de Estructura de la Materia. 
Consejo Superior de Investigaciones Cient{\'\i}ficas. 
Serrano 123, 28006 Madrid. Spain. \\
$^{\ddag}$
Instituto de Ciencia de Materiales.
Consejo Superior de Investigaciones Cient{\'\i}ficas.
Cantoblanco. 28049 Madrid. Spain.  \\
$^*$
Departamento de Matem\'aticas. Universidad Carlos III.
Butarque 15. Legan\'es. 28913 Madrid. Spain.}
\date{\today}
\maketitle 
 
\begin{abstract}

The influence of electron-electron scattering on quasiparticle 
lifetimes in graphite is calculated.
In the limit when the
Fermi surface is reduced to isolated points in the Brillouin Zone,
the suppression of screening 
leads to non Fermi liquid behavior. The inverse lifetime increases linearly
with energy, in agreement with recent experiments.
Similar features should also be present in
narrow gap semiconductors, and in carbon nanotubes.

\end{abstract}


%\narrowtext

Recent experiments\cite{exp} show that the inverse 
lifetime of quasiparticles in
graphite increases linearly with energy over a broad energy range
$\sim 0.3 - 4$eV.  The authors of ref.\cite{exp} argue that this effect
may be due to the unusual dispersion relation of plasmons in
a 3D stack of conducting planes, where the electrons
are confined to move in two dimensions\cite{plasmon}. In the present work,
we analyze an alternative explanation of the energy dependence
of the lifetime, although also associated to the electron-electron
interaction. 

Graphite is a semimetal. Band structure\cite{band} calculations show that
electrons are confined to narrow holes near the edges of the
hexagonal Brillouin zone. The main features around the
Fermi level are well described 
by a simple H\"uckel theory\cite{Hueckel}, which includes
only the out of plane $\pi$ orbitals located at each 
carbon atom. Intraplane hopping is much larger than the interplane
hybridization. If we neglect the coupling between planes, the
Fermi surface is reduced to two isolated points at the corners
of the 2D Brillouin zone. The band dispersion is linear around
these points, $\epsilon_k = \hbar v_F | \vec{k} |$. In units
of the hybridization between neighboring $\pi$ orbitals, $t$,
we find that $\hbar v_F = 3 t a /2 $ where $a$ is the distance between
carbon atoms. The density of states is zero at
the Fermi level, within this approximation. The long wavelength
properties can be approximated by means of an effective 2D
Dirac equation\cite{Dirac}, instead of the more familiar
effective mass approximation, used when the bands are parabolic
at low energies.

Because of the vanishing density of states at the Fermi energy,
the Coulomb interaction is not screened in a conventional fashion.
Combining the band structure described above with the bare
Coulomb interaction we obtain an effective hamiltonian,
for a single plane of graphite:

\begin{equation}
{\cal H} = \hbar v_F \int d^2 r \bar{\Psi} ( \vec{r} )
( i \sigma_x \partial_x + i \sigma_y \partial_y )
\Psi ( \vec{r} ) + \frac{e^2}{2 \epsilon_0}
\int d^2 r_1 \int d^2 r_2 \frac{\bar{\Psi} ( \vec{r}_1 )
\Psi ( \vec{r}_1 ) \bar{\Psi} ( \vec{r}_2 ) \Psi ( \vec{r}_2 )}
{| \vec{r}_1 - \vec{r}_2 |}
\label{hamil}
\end{equation}

where $\sigma_x$ and $\sigma_y$ are Pauli matrices,
and $\Psi$ is a two component spinor associated
to the two bands which meet at the Fermi points. In
a honeycomb carbon lattice, $t \approx 2.4$ eV.
High energy screening
processes are included in the effective dielectric constant,
$\epsilon_0$.
We expect that (\ref{hamil}) describes well the physics of a single
layer of graphite for energies $\le t$. 

The hamiltonian (\ref{hamil}) defines a renormalizable theory\cite{graphite}
in the field theoretical sense\cite{Shankar,Polchinski,book}. 
Perturbation theory 
leads to logarithmic dependences on the high energy cutoff,
which can be incorporated into new, renormalized, parameters
in a standard way.  The existence of logarithmic 
corrections can be inferred from the fact that
the actual coupling constant in (\ref{hamil}),
$e^2 / ( \epsilon_0 \hbar v_F )$ is dimensionless.

A detailed analysis of the renormalization procedure used to
deal with the 2D hamiltonian (\ref{hamil}) is given in ref.\cite{graphite}.
That scheme can be generalized to a system
of weakly interacting planes.  Interplane couplings can be of two types:

i) Coulomb effects. They increase the number of diagrams that
need to be calculated, as an electron in a given plane can be
scattered by electron-hole pairs in other planes. The low energy
properties are, however, not affected. The RG scheme proceeds by
integrating out slices in energy and momentum space. The momentum
transfer  in interplane processes 
is bounded by the inverse of the interlayer
distance, $d^{-1}$. Thus, at sufficiently low scales, the
electron-electron scattering retains the original dependence on
the momentum transfer. The finite corrections induced by interplane
Coulomb interactions can be included in a manner similar to
the scheme used in ref.\cite{plasmon}. The details are given below.

ii) Interplane electron hopping. This interaction is responsible
for the 3D features of the bands of graphite.  It leads to deviations
from the linear dispersion relation used to define (\ref{hamil})
at low energies. We expect that these deviations will become significant
at energies comparable to the interplane hopping, $\approx 0.27$
eV\cite{band}. In addition, the density of states at the Fermi level
becomes finite.
Hence, metallic screening takes place at length scales greater
than $k_{FT}^{-1}$, where $k_{FT}^2 = 4 \pi e^2 N ( \epsilon_F )$.
Because of the smallness
of $N ( \epsilon_F )$\cite{band}, the associated energy scale, $\hbar v_F
k_{FT}$ is much smaller than the previous one.

Quasiparticle lifetimes are also influenced by phonons\cite{phonon},
and by low energy, out of plane plasmons\cite{plasmonz}. 
The phonon bandwidth in graphite is $\sim 0.20$ eV, and 
the out of plane plasmons have energies $\sim 0.05$ eV.
Thus, for quasiparticle energies $> 0.2$ eV, these processes
should give a constant contribution, independent
of the quasiparticle energy.

The preceding discussion implies that a straightforward generalization of the 
hamiltonian (\ref{hamil}), including the existence of
a stack of planes coupled solely by the Coulomb interaction
describes well the electronic properties of graphite, in the 
energy range $0.2$ eV $ < \epsilon < 2$ eV. 

A remarkable feature of the perturbation analysis of hamiltonian
(\ref{hamil}) is that logarithmic divergences appear in
the corrections to one particle properties, like the self energy,
but electron-hole propagators are finite\cite{graphite}. 
This reflects the fact that the divergences are due to the singularity
of the interaction, and not to density of states effects.
The intra BZ edge electron-hole propagator 
at low energies and momenta is\cite{graphite}:

\begin{equation}
\chi_0 ( \omega , \vec{q} ) = \frac{\vec{q}^2}{32 \pi \sqrt{
v_F^2 \vec{q}^2 - \omega^2}}
\label{susc}
\end{equation}

where $\chi_0$ is purely real for $v_F | \vec{q} | > \omega$,
and purely imaginary otherwise. Thus, electron-hole pairs can only be
excited if $v_F | \vec{q} | < \omega$. This region is shown 
in fig. (\ref{fsusc}).

The screened Coulomb potential, including interplane scattering,
can be written as\cite{plasmon}:

\begin{equation}
v_{scr} ( \omega , \vec{q} ) =
\frac{2 \pi e^2}{ \epsilon_0 | \vec{q} | } \frac{\sinh ( | \vec{q} | d )}
{\sqrt{ \left[ \cosh ( | \vec{q} | d ) + \frac{2 \pi e^2}
{ \epsilon_0 | \vec{q} |}
\sinh ( | \vec{q} | d )  \chi_0 ( \omega , \vec{q} ) \right]^2 - 1 }}
\label{suscrpa}
\end{equation}

where $d$ is the distance between planes.
It is interesting to
note that ${\rm Re} \ v_{scr}$ has no poles at low energies,
unlike for the case of a stack of layers with  quadratic  dispersion,
where a plasmon band,
$\omega_p \propto | \vec{q} |$, was found\cite{plasmon}.
We adscribe this difference to the fact that,
in the present case, electron-hole pairs can only exist
for $\omega > v_F | \vec{q} |$, while in a conventional
electron gas it is the opposite.

Using (\ref{suscrpa}), the inverse of the quasiparticle lifetime
can be written as:

\begin{equation}
{\rm Im}  \Sigma ( \omega , \vec{k} ) =
\frac{2}{4 \pi^2} \int d^2 k' \frac{1 + \cos ( \phi_{\vec{k} -
\vec{k'}})}{2} 
{\rm Im} \ v_{scr} \left(
\omega - \epsilon_{k'} ,  \vec{k} - \vec{k'} \right)
\label{sener}
\end{equation}

where $\phi_{\vec{k} - \vec{k'}}$ is the angle between vectors 
$\vec{k}$ and $\vec{k'}$, and we are summing over the two spins.

The expression (\ref{sener}) can be interpreted as the probability for
a quasiparticle with frequency $\omega$ and momentum $\vec{k}$
to decay into a real quasiparticle of energy $\epsilon_{k'}$
and momentum $\vec{k'}$. Kinematical constraints in the phase space
of final states imply that ${\rm Im} \Sigma ( \omega ,
\vec{k} ) \ne 0$ only if $\omega \le v_F | \vec{k} |$.
This restriction seems incompatible with the
phase space available for the creation of electron-hole
pairs, shown in fig.(\ref{fsusc}), suggesting that there are no channels
for quasiparticle decay in the model described by (\ref{hamil}).

We must, however, consider with care the limit
$\lim_{\omega \rightarrow \epsilon_k + 0^+}  {\rm Im}
\Sigma ( \omega , \vec{k} )$. The simplest situation, which can be 
analyzed analytically, is the lowest order diagram shown
in figure (\ref{fsener}).  For this case, 
${\rm Im} \Sigma ( \omega , \vec{k} )$
drops discontinuously to zero at $\omega = v_F | \vec{k} |$.
The magnitude of the step is:

\begin{equation}
\lim_{\omega \rightarrow \epsilon_k + 0^+}
{\rm Im} \Sigma ( \omega , \vec{k} ) = 
\frac{1}{48} \left( \frac{e^2}
{\epsilon_0 \hbar v_F} \right)^2  \hbar v_F | \vec{k} |
\label{ltime}
\end{equation}

The inverse lifetime, defined in this way,  increases linearly
with the energy of the quasiparticles. The real part of the
self energy shows a logarithmic dependence on the high energy cutoff 
needed to define the model (\ref{hamil}), leading to
non Fermi liquid behavior\cite{graphite}. The existence of a finite
lifetime, despite
the kinematical constraints described earlier, can be traced back to 
the divergence of the density of electron-hole pairs
in the forward direction, $\omega = v_F | \vec{k} |$, which
compensates exactly the reduction in the number
of states in which the quasiparticle can decay as
$\omega \rightarrow \epsilon_k$. Setting $\hbar v_F = 3/2 t a$,
where $t = 2.6$ eV, $a = 1.4$ \AA  and $\epsilon_0 = 2.4$\cite{diel},
we find that the constant of proportionality between the
inverse lifetime and the quasiparticle energy is
0.049 in eV$^{-1}$ fs$^{-1}$. This value compares well to the
experimental one, 0.029 eV$^{-1}$ fs$^{-1}$\cite{exp}. 
Note that the inverse lifetime should be an average
of ${\rm Im} \Sigma$ over a finite interval of
energies and momenta (see below), and expression (\ref{ltime})
is an upper bound to such an average.

The behavior of 
${\rm Im} \Sigma ( \omega , \vec{k} )$ as
$\omega \rightarrow \epsilon_k$,
including the RPA and interplane interactions, 
can be calculated numerically, and it is shown in figure (\ref{flcone}),
using expression (\ref{sener}), with $d = 3.35$ \AA and
the parameters given above.

The kinematical constraints discussed earlier arise from 
the requirement of momentum conservation. In the presence of
disorder, quasiparticles have a finite spread in momenta.
Because of the sharp rise of ${\rm Im} \Sigma$ away from the
line $\omega = \epsilon_k$, this spread leads also to a
finite quasiparticle lifetime. This is shown in
fig. (\ref{fltime}), where ${\rm Im} \Sigma ( \epsilon_k ,
| \vec{k} |  - \Delta k )$ is plotted, with 
$\Delta k = 0.002$ \AA$^{-1}$.
$( \Delta k )^{-1}$ corresponds, roughly,
to the mean distance beween scattering centers,
in our case, $l \sim 500$ \AA. Other inelastic 
scattering channels, such as phonons, will also contribute to 
give a spread in momentum and energy to the
quasiparticles\cite{phonon}.

The lifetimes shown in fig. (\ref{fltime}) are consistent with the
experimental observations\cite{exp}. The explanation that we propose
here also implies non Fermi liquid behavior in other properties of
graphite, such as the conductivity or the susceptibility. 
Note, however, that we expect our model to break down at low
energies, $< 0.2$ eV.

It was argued that low energy plasmons can be responsible
for the unconventional quasiparticle lifetimes in graphite\cite{exp}.
A layered 2D electron gas has plasmons above a certain threshold
which depends linearly on momentum, $\omega_{treshold} = v_{pl} 
| \vec{q} |$\cite{plasmon}. These collective excitations give rise
to a new decay channel for quasiparticles with velocity
$v_{qp} = \hbar k / m > v_{pl}$. In most cases, 
decay into low energy plasmons cannot take place at the Fermi
level, because $v_{pl} > v_F$. It is proposed in ref\cite{exp}
that, in graphite, $v_{pl} < v_F$.  In that case, the low energy
plasmons
merge into the electron-hole continuum. The influence of these
hybridized excitations on the lifetimes is rather complicated,
and does not lead to a precise linear dependence on the energy of
the quasiparticles. We believe that, in any case, the peculiar band
structure of graphite implies that screening is dominated
by interband transitions, which preclude the existence
of in plane low energy plasmons, as deduced from eq.(\ref{susc}).

The results depicted in fig.(\ref{fltime}) have been
obtained by combining a many loop propagator, the RPA bubble
modified by interplane effects, with a zero loop
description of the quasiparticles. We neglect the renormalization
of the quasiparticle pole, calculated in \cite{graphite}.
As we scale towards low energies, the quasiparticle pole loses
spectral strength, in the manner analyzed in \cite{graphite}.
We expect, however, the main result of this paper, 
shown in fig. (\ref{fltime}), to be weakly dependent on 
this renormalization. Unlike in 1D conductors, the hamiltonian
(\ref{hamil}) flows towards a free fixed point\cite{graphite},
which makes plausible the use of unrenormalized quasiparticle
propagators in the calculation of the lifetimes.
Note, however, that the bare coupling constant, $e^2 / \epsilon_0
\hbar v_F$, is of order unity, although most likely reduced by
density of states factors, $ \sim ( 2 \pi )^{-2}$.

The main physical basis for the unconventional dependence of
the quasiparticle lifetimes in energy, and the breakdown of
Fermi liquid theory, lies in the absence of metallic screening.
Thus, we expect that other semimetals may exhibit similar behavior.
In particular, it is well known that the band structure of zero
gap semiconductors can be approximated by the 3D Dirac
equation\cite{zerogap}. In these materials, we expect that a
description similar to the one used here to be valid
down to the lowest energies.
The graphite structure is the basis
of the carbon nanotubes\cite{nanotube}, which should also exhibit
the non Fermi liquid behavior analyzed here.

\begin{figure}
\caption{Region in phase space available for electron-hole
excitations. The nature of the interband e-h pairs is
sketched in the diagram.}
\label{fsusc}
\end{figure}

\begin{figure}
\caption{Lowest order contribution to the quasiparticle lifetime.}
\label{fsener}
\end{figure}

\begin{figure}
\caption{Imaginary part of the selfenergy as function of frequency for
various momenta: Full line, $k = 0.1$ \AA$^{-1}$, dashed line,
$k = 0.2$ \AA$^{-1}$, broken line, $k = 0.4$ \AA$^{-1}$.}
\label{flcone}
\end{figure}

\begin{figure}
\caption{Inverse quasiparticle lifetime, as defined in the text,
as fucntion of the energy of the quasiparticle.}
\label{fltime}
\end{figure}

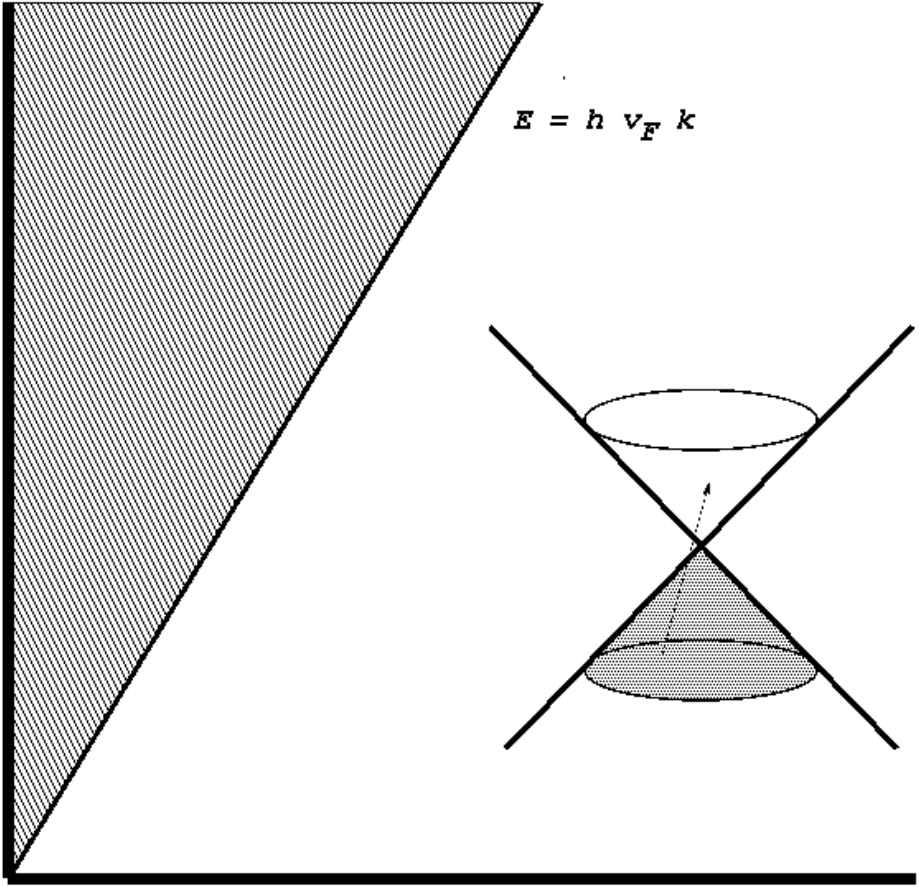
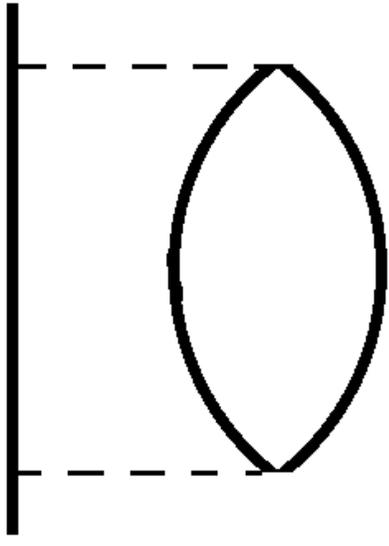
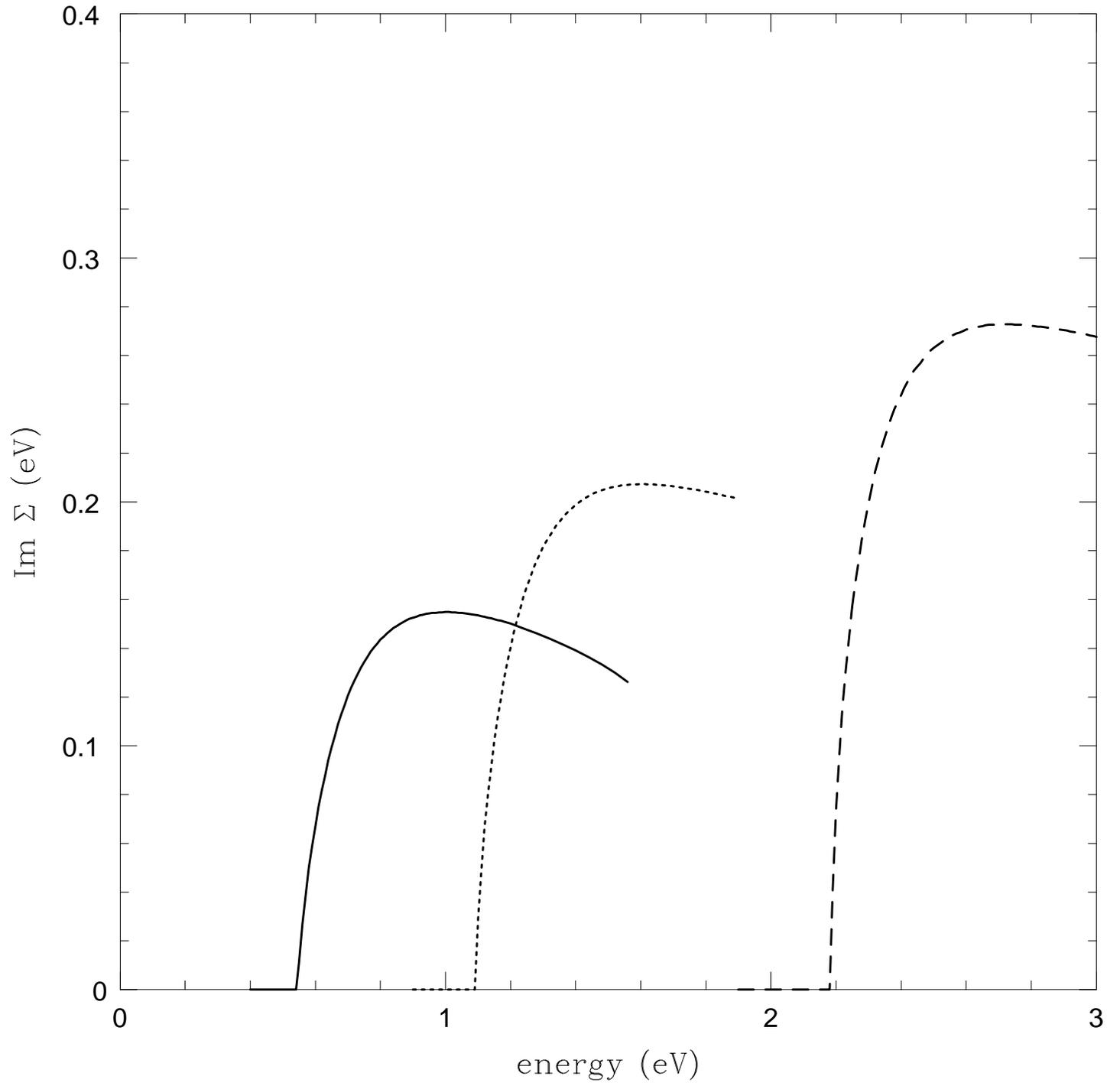
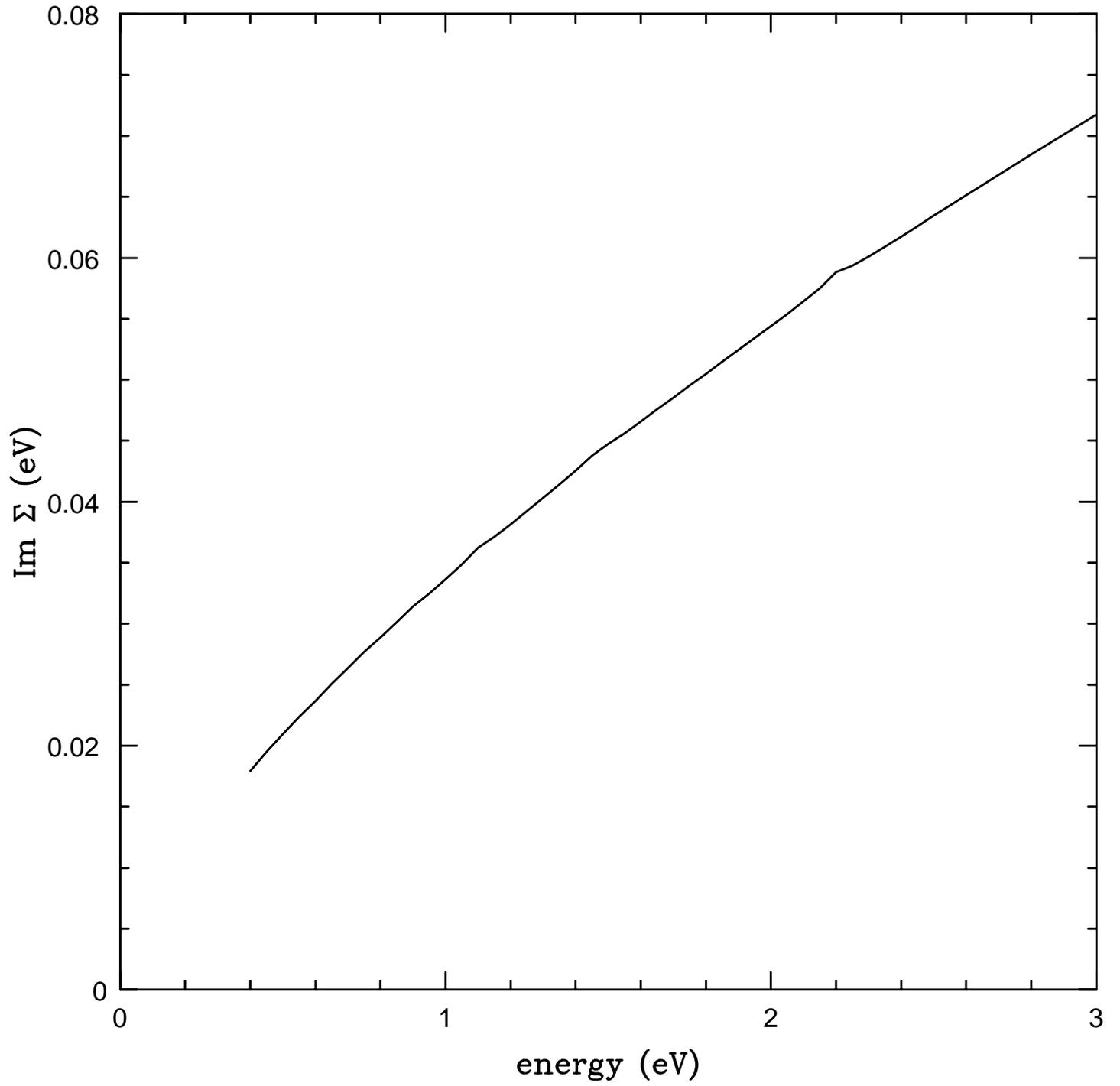
\end{document}